\documentclass[10pt,draft]{dis03}
\usepackage{amsmath}
\usepackage{epsfig}
\usepackage{amssymb}

\begin{document}

\title{Parton energy loss or nuclear absorption: \\ What quenches hadron spectra at HERA ?\thanks{Proceedings XI International Workshop on Deep Inelastic Scattering (DIS03), St. Petersburg, 23-27 April 2003}}

\author{Fran\c{c}ois Arleo \\
ECT* and INFN, G.C. di Trento \\
Strada delle Tabarelle, 286\\
38050 Villazzano (Trento), Italy \\
E-mail: \texttt{francois@ect.it}}

\maketitle

\begin{abstract}
\noindent We suggest and explore two observables which may clarify the origin of the attenuation of semi-inclusive hadron production reported in DIS on nuclear targets.
\end{abstract}

\section{Introduction} 

A significant depletion of semi-inclusive hadron production in Deep Inelastic Scattering (DIS) on nuclear targets has been reported by the HERMES collaboration. The attenuation of various hadron species ($\pi^\pm$, $K^\pm$, $p$, $\bar{p}$) on helium, nitrogen, and krypton nuclei now extends to a wider kinematic range~\cite{Airapetian:2003mi}, while mesurements on neon targets were presented for the very first time at this workshop~\cite{Elbakyan}.

Despite this impressive high-statistics data set, however, the origin of the observed quenching remains unclear. On the one hand, final state inelastic interaction of the produced hadron (or pre-hadron) in the nuclear medium may be responsible for the measured attenuation~\cite{Accardi:2002tvFalter:2003di}. On the other hand, it has later been argued that the multiple scattering process incurred by hard quarks and gluons in dense media could affect the fragmentation mechanism and subsequently deplete semi-inclusive hadron production on nuclei~\cite{Wang:2002riArleo:2002kh}.

The gross features of the present data, such as the $\nu$ or $z$ dependence (respectively the photon energy and the momentum fraction carried away by the leading hadron) of the hadron attenuation, do not yet allow for a clear discrimination between those two competing processes. It is the aim of these proceedings to discuss two observable quantities which we hope could clarify to some extent the origin of the reported nuclear effects. More precisely, we suggest (i) to compare semi-inclusive hadron production on two large nuclei (Section~\ref{se:largenuclei}), and (ii) to look for the $Q^2$ dependence of isospin effects (Section~\ref{se:Q2dependence}).

Both observables will be predicted below assuming the sole effect of parton energy loss, neglecting any subsequent hadronic final state interaction in the medium. The details of our present calculations being not reproduced here due to the lack of space, we refer the reader to our recent analysis Ref.~\cite{Arleo:2003jz} for completeness.

\section{Comparing large nuclei}\label{se:largenuclei}

The parton energy loss is directly linked to the transverse momentum 
\begin{equation}\label{eq:omc}
\omega_c \,=\, \frac{1}{2}\,\hat{q}\,L^2
\end{equation}
accumulated by the soft gluons radiated by the hard parton all along its way through the dense QCD medium~\cite{Baier:2000mf}. The transport coefficient $\hat{q}$, which measures the density of scattering centers in the medium, is here determined from Drell-Yan production data~\cite{Arleo:2002ph}. The variable  $L$ in~(\ref{eq:omc}) stands for the hard parton in-medium pathlength and depends on two scales, namely, the nuclear radius, $R$, and the hadron formation time, $t_f$. Assuming for simplicity a sharp sphere nuclear density profile, it reads~\cite{Arleo:2003jz}
\begin{eqnarray}\label{eq:meanl}
L &=& t_f\,\times\,\left[ 1 - \frac{3}{8}\,\frac{t_f}{R} + \frac{1}{64}\,\left(\frac{t_f}{R}\right)^3\right] \hspace{1.05cm} \mathrm{if} \, \, t_f \leq 2\,R \nonumber \\
L &=& \frac{3}{4}\, R, \hspace{5.cm}  \mathrm{if} \,\,t_f > 2\,R
\end{eqnarray}
When the hadron formation time is large as compared to the nuclear radius, the energy scale $\omega_c$ (hence, the parton energy loss) thus depends quadratically on the nuclear size. In this limit (``small'' nuclei), we expect a rather strong dependence of the parton energy loss on the atomic mass number, $A^{2/3}$. Conversely, for hard quarks produced in ``large'' nuclei (i.e. for which $R \gg t_f$), the covered length $L \simeq t_f$ is only given by the hadron formation time. As a consequence,  the quenching of semi-inclusive hadron spectra becomes independent of the nuclear size $R$ (or $A$). 

An opposite behavior is expected if hadronic absorption is the dominant process at work: the larger the nucleus, the stronger the depletion of hadron spectra. Therefore, we believe that comparing semi-inclusive hadron production on two large nuclei $A$ and $B$ with radii respectively $R_A \gtrsim R_B \gg t_f$ should unravel the underlying mechanism.

\vspace*{6.6cm}
\begin{figure}[h]
\begin{center}
\includegraphics{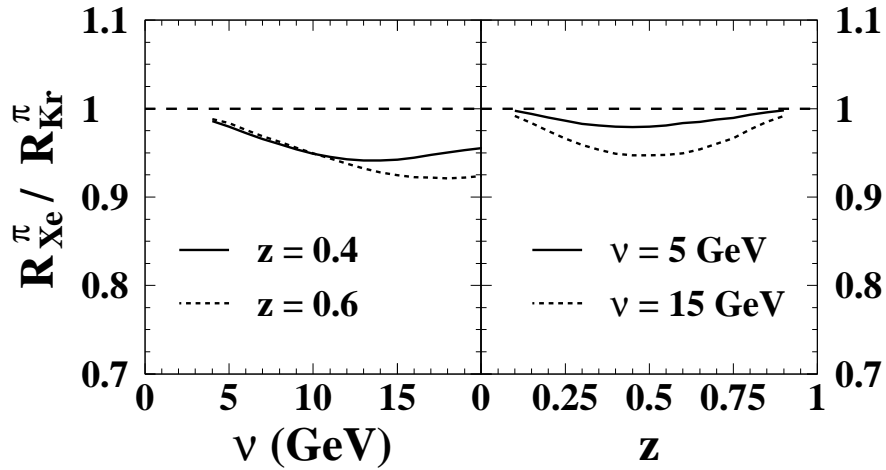}
\caption{Ratio of semi-inclusive hadron production on xenon over krypton targets as a function of $\nu$ ({\it left}) and $z$ ({\it right}). Calculations are detailed in Ref.~\cite{Arleo:2003jz}.}
\label{fig:pbkr}
\end{center}
\end{figure}

\vspace*{-1.cm}
To illustrate this, the semi-inclusive charged pion production ratio on xenon over krypton targets is computed in Figure~\ref{fig:pbkr} as a function of $\nu$ ({\it left}) and $z$ ({\it right}) for various $z$ and $\nu$ bins respectively. Although the atomic mass numbers are quite different ($A = 84$ and 132), the ratio is remarkably close to one. It would then be most interesting to contrast this result to the predictions based on hadronic absorption models.

\section{$Q^2$ dependence of isospin effect}\label{se:Q2dependence}

A striking feature of the recent HERMES results is the noticeable isopin effect reported in the kaon sector, i.e. the stronger depletion for negative than for positive kaons, $R^{K^-}(\nu) < R^{K^+}(\nu)$. Such a behavior is also observed in our theoretical predictions based on parton energy loss\footnote{Note however that hadronic absorption models also account for this behavior from the stronger $K^-$ inelastic interaction with the surrounding medium.}. When Bjorken $x = Q^2 \,/\, 2M\nu$ is not too small, virtual photons mostly couple to valence up quarks ($u(x \lesssim 1) \gg \bar{u}(x)$) which preferably fragment into $(u\bar{q})$ (valence-type channel) than into $(\bar{u}q)$ (sea-type) states. This translates into a steeper slope for the $z$ dependence of $u\to K^-$ fragmentation functions, hence into a stronger attenuation in this channel~\cite{Arleo:2003jz}. 

\vspace*{7.4cm}
\begin{figure}[h]
\begin{center}
\includegraphics{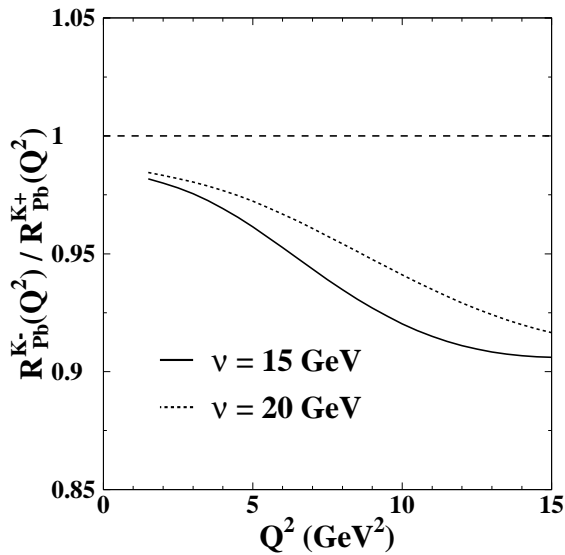}
\caption[*]{Ratio of $K^-$ over $K^+$ attenuation on a lead nucleus as a function $Q^2$ for $\nu = 15$~GeV and $\nu = 20$~GeV. The momentum fraction is set to $z = 0.6$.}
\label{fig:q2dep}
\end{center}
\end{figure}

\vspace{-0.8cm}
At very small $x$, nevertheless, the fragmentation of sea quarks --~for which $u(x \ll 1) \simeq \bar{u}(x)$~-- becomes the dominant channel in semi-inclusive hadron production. Subsequently, no more isospin effect is expected, that is $R^{K^-}\simeq R^{K^+}$. This could be observed experimentally through the $x$ dependence (or equivalently the $Q^2$ dependence at a given $\nu$) of the $K^-$ over $K^+$ attenuation ratio. Our predictions of the kaon quenching ratio on a lead nucleus at two different energy ($\nu = 15$ and 20~GeV) are shown in Figure~\ref{fig:q2dep}. The smoothly decreasing ratio with $Q^2$ clearly indicates the transition from small to not too small Bjorken $x$, where isospin effects get amplified. We believe that such a behavior  would therefore be a clear hint that the nuclear medium acts on semi-inclusive hadron production at a partonic level.

\section{Summary}

Two opposite scenarios have been proposed to account for the observed attenuation of semi-inclusive hadron spectra in DIS on nuclei: hadronic absorption or partonic multiple scattering. It is therefore a challenging issue to disentangle --~and quantify~-- these two processes. In these proceedings, we explored two observable quantities which appear promising to reach this goal:
\begin{enumerate}
\item In ``large'' nuclei, the propagating degrees of freedom are essentially hadrons. With in this respect, the production ratio on two large targets predicted by these alternatives should be rather different;
\item The isospin effect measured in semi-inclusive kaon production may prove a useful tool. We stressed in particular that the $Q^2$ dependence of the $K^-$ over $K^+$ suppression is a sensitive probe of parton multiple scattering in the nuclear medium.
\end{enumerate}

\end{document}